\documentclass[12pt]{article}
\usepackage{amsmath, amscd, amssymb, latexsym, epsfig, color, amsthm,enumerate, mathtools}
\usepackage[all]{xy}
\usepackage{verbatim}
\usepackage{graphicx}
\usepackage{amsmath}
\usepackage{float}

\newcommand{\beq}[1]{\begin{equation}\label{#1}}
\newcommand{\enq}[0]{\end{equation}}

\newcommand{\remove}[1]{}

\title{Conjecture C Still Stands}
\author{Gil Kalai\thanks{Einstein Institute of Mathematics, Federmann Center for the Study of Rationality,
    and Quantum Information Science Center, The Hebrew University of Jerusalem,
  and Efi Arazi School of Computer Science, and Center of Foundations and Applications of Cryptography,
  Reichman University, Herzliya.}
}
\begin{document}
        \maketitle

        \begin{abstract}
          
          More than ten years ago the author
          described a parameter $K(\rho )$ for the complexity of $n$-qubit quantum state $\rho$
          and raised the conjecture (referred to as ``Conjecture C'')
          that when this parameter is superpolynomial in $n$, the state $\rho$
          is not experimentally feasible (and will not be experimentally achieved without
          quantum fault-tolerance). Shortly afterward \cite{FlaHar13},
          Steve Flammia and Aram Harrow claimed that
          the simple easy-to-construct
          $W$ states are counterexamples to
          ``Conjecture C.'' 
          We point out that Flammia and Harrow's argument regarding $W$-states is incomplete.
          Moreover, the emergent picture from experimental progress of the past decade on
          noisy intermediate scale quantum (NISQ) computers 
          suggests
          that $W$-states, as simple as they appear,
          cannot be achieved experimentally by NISQ computers,
          and can not be constructed without quantum fault-tolerance.

\end {abstract}
        \section {Sure/Shor separators}

        Let $\rho$ be a quantum state on the space $(\mathbb C^2)^{\otimes n}$ of $n$-qubits.
        We denote the set of qubits
        by $[n]=\{1,2,\dots, n\}$. Kalai \cite {Kal:d,FlaHar13} defined a function $K(\rho)$,
        which, in a sense, measures the complexity of $\rho$ in terms of entanglement.

        The definition of $K(\rho)$ requires a
        few steps as follows.
        First, for a subset $S$ of qubits denote by $\rho_S$ the reduced state of $\rho$
        onto $S$. Next, partition $S$ further
into two nonempty subsets of qubits $A$ and $B:= S \backslash A$ (thus $S$ must contain at least two qubits).
Denote by $Sep(A)$ the convex set of quantum states that are bipartite separable on $S$ across the
bipartition defined by $A$ and $B$. Next, define
$$\Delta (\rho_S)= \min _{A \subset S} \inf _{\sigma \in Sep (A)} \|\rho_S-\sigma\|_{tr}.$$
In words, $\Delta (\rho_S )$ measures the distance in the trace norm between the nearest biseparable state and $\rho_S$,
across all bipartitions of $S$. 
Now define 

        \begin {equation}
          K(\rho)=\sum _{S \subset [n]} \Delta (\rho_S).
        \end {equation}

        In the 2012 debate between Gil Kalai and Aram Harrow,
        Kalai \cite {Kal:d} made the following conjecture.

\bigskip

{\bf Conjecture C:} For all states $\rho$  that have been prepared by a noisy quantum
computer, there is a polynomial $P(n)$ such that $$K(\rho ) \le P(n).$$

\medskip


Flammia and Harrow \cite {FlaHar13} proposed two counterexamples
to Conjecture C and described the purpose of the conjecture
in the following words

\begin {quotation}
\noindent
``Conjecture C, as we see it, is an attempt at a `Sure/Shor separator' (as Scott Aaronson puts
it \cite {Aar04}) that would distinguish states we have definitely already seen from the sort of states we
would find in a quantum computer. It represents an admirable attempt to formulate quantum-computing
skepticism in a rigorous and testable way.''
\end {quotation}

We can also think about Sure/Shor
separators as a distinction between quantum states that can be described
by current and near-future noisy quantum computers, and
quantum states whose preparation would require quantum fault-tolerance.
Conjecture C (and likewise 
some earlier proposals for Sure/Shor separators
from \cite {Aar04} and \cite {Kal09}) is of an asymptotic nature and proposes
a distinction between ``theoretically possible'' and ``empirically witnessed'' states
when the number of qubits grows.

We note that 
in later works \cite {Kal16,Kal18,Kal20} (following \cite {KalKin14}) Kalai put forward an
argument that distance-5 surface codes
(and other good-quality quantum error-correcting codes)
are themselves (inherently) ``Sure/Shor separators.''
(This argument was put forward a few years after the Kalai--Harrow debate and \cite {FlaHar13}.)   
Of course, if one accepts that good-quality quantum error-correcting codes such as distance-5
surface codes are themselves
Sure/Shor separators and cannot be achieved without quantum fault-tolerance,
then, since those states are necessary ingredients for quantum fault-tolerance,
it would follow that quantum fault-tolerance itself must be out of reach.\footnote{In fact, it would follow
  that David DiVincenzo's step four
  in his famous seven-step road map to quantum computers \cite {DiV00} is already out of reach.}

Flammia and Harrow first pointed out that Conjecture C cannot be extended to qudits rather than qubits.
Starting with a very simple state $\rho$ on $16n$ qubits (where $K(\rho)$ is linear in $n$), by grouping
the qubits into groups of 4 one gets an exponential value for $K_d(\rho)$. ($K_d$ is the function obtained by
applying the definition of $K(\rho)$ to qudits.)
This is a correct observation, albeit not a counterexample
to the conjecture as Kalai originally proposed.\footnote
{We can take $4n$ pairs of qubits in cat states ($n=3m$), then divide the qubits into $2m$ triples of qubits at
  random, and regard each
  triple as a single qudit. This will artificially create a complicated entanglement
  structure between the qudits that disappears when we
  consider the original qubit tensor-power structure.} 
Next Flammia and Harrow considered the familiar state $W_n$, showed that $K(W_n)$
is exponential in $n$, and claimed that
the states $W_n$ are counterexamples to Conjecture C because the physical plausibility of
these states is ``relatively uncontroversial.''

In this note we point out that Flammia and Harrow's argument
regarding $W$-states is incorrect or at least incomplete.
The emergent picture from experimental progress of the past decade on
          noisy intermediate scale quantum (NISQ) computers 
          suggests
          that $W$-states, as simple as they appear,
          currently cannot be achieved experimentally by NISQ computers, 
          and may not be constructed without quantum fault-tolerance.
          The proposed realization of $W$-states in natural quantum systems are interesting but
          an argument for why the parameter $K(\rho)$ remains exponential is missing, and the
          situation for NISQ computer may shed doubt if this is at all true.

          We also point out in Section \ref {s:d} some fallacies in Steve and Aram's approach
          that may have a bearing on the larger debate on quantum computers.
          
\section {The $W$-states}

Define the state $W_n$ on $n$ qubits as follows:
        \begin {equation}
  \left | W_n \right\rangle = \frac {1}{\sqrt n} \sum_{k=1}^{n}  \left | 0^{k-1},1,0^{n-k} \right\rangle.
  \end {equation}

\noindent
Flammia and Harrow showed that $$K(W_n) \ge 2^{n-4}.$$

{\bf Remark:} This observation by Flammia and Harrow was indeed useful to our ongoing 2012 debate and
it took me by surprise at the time. Previously
I had expected that the exponential behavior of
$K(\rho)$ would genuinely represent complicated quantum states.
In 2012 I did not have reason to doubt the claim by Flammia and Harrow that
$W_n$ are experimentally feasible. 

The states $W_n$ are indeed simple, but can these states be constructed?
Do they represent ``previously observed physics?''
One avenue proposed in \cite {FlaHar13} to prepare $W_n$ is as follows:

\begin {quotation}
\noindent  
``One appealing feature of the state $\left |W_n \right \rangle$
is that it can
be prepared in the electronic spin states of some kinds of atoms \cite {SvHM04}.
For example, we can prepare
it by first polarizing the atoms, putting them in the $\left |0^n \right \rangle$
state, and then sending in a single longwavelength photon. If this photon is absorbed,
it will cause exactly one excitation,
and if the
wavelength of the photon is long enough, this excitation will be completely delocalized among all
the atoms. We remark that some trapped ion experiments have already
prepared the state $\left |W_8 \right \rangle$ 
with reasonably high fidelity \cite {Hai05}.''\footnote{A related proposal by
Greg Kuperberg (private communication, 2022) is to consider as a realistic representation of $W_n$ (where $n$
can be very large) the state of a single electron delocalized among the $n$ atoms \cite {Edw13}.
Kuperberg expects that a cold, clean gold wire would give a $W_n$-state (which violates Conjecture C
for a natural qubit structure)
for $n=10^9$.}  

\end {quotation}


Here Flammia and Harrow attempt to demonstrate the state $W_n$ as a state occurring in nature
with respect to a certain qubit structure. This represents a wider yet welcome interpretation
of conjecture C. (It is a wider interpretation since referring to a quantum computer
usually assumes not only a qubit structure but also the
ability to perform or witness gates operating on these qubits.) 
Of course, even if one can demonstrate $W_n$-like states for large values of $n$, a remaining
question would be if the deviation from ideal $W_n$ 
suffices to destroy the exponential value of $K(W_n)$.
The proposed realization of $W$-states by natural quantum systems by Flammia and Harrow
are interesting and they were useful for advancing the 2012 debate \cite {Kal:d},
but an argument for why the parameter $K(\rho)$ remains exponential is missing and so is a study of 
noise models for the proposed realization.

Another  avenue to study Conjecture C in its original form
is to use noisy intermediate-scale quantum (NISQ) computers to  build the
state $\left |W_n \right \rangle$. It is not even known that
bounded depth circuits suffice to create $W_n$ (see \cite {MM}) and even for
such circuits the fidelity decreases exponentially with the number of qubits.
Based on what we know about NISQ systems, we
cannot expect them to lead to high-fidelity $W_n$ states
(and probably not even low-fidelity $W_n$ states $\psi$ with
a large value of $K(\psi)$).

All in all, when we look at the experimental progress in the past decade,
and especially the difficulty to control quantum computers with a handful of qubits,
it appears that the claims
in \cite {FlaHar13} that $W_n$ represent ``previously observed physics'' are not supported.

\section {Discussion}
\label {s:d}

The short paper of Flammia and Harrow contains an
interesting discussion that, in my opinion, concisely brings together
some prevalent fallacies regarding quantum computers. The points raised in this section
are also relevant to some critique of Eric Aurel
on my papers (\cite {Aur18}), and to the debate
on quantum computing as a whole. 


\subsection* {1. The fallacy of ``harnessing quantum physics''}

Flammia and Harrow write:

\begin {quotation}
\noindent  
``However, we believe that our counterexamples are significant not especially because they refute
Conjecture C, but because they do so while side-stepping Kalai's main points about quantum error
correction failing. More generally, it is significant that it is so hard to come up with a sensible
version of Conjecture C. In our view, this is because quantum computers harness phenomena, such
as entanglement and interference, that are already ubiquitous, but merely hard to control.''

\end {quotation}

In my view, the fallacy expressed in this quote is significant and widely accepted;
many researchers regard complicated quantum states
of the kind promised by the quantum computer model are already ubiquitous but ``merely hard to control.''
This belief is sometimes
carried over to the noisy states created by present-day NISQ computers,
as people sometimes think that the added noise only makes those
states ``computationally even more complicated while harder to control.''

\subsection* {2. The fallacy of using thought experiments as empirical evidence}

Here are two more quotes from \cite {FlaHar13}:

\begin {quotation} 
\noindent  
``The goal of this note is to create the simplest possible thought
experiment to refute Kalai's Conjecture C, but variants of that conjecture could be countered with
variants of this thought experiment.''

\end {quotation}

\begin {quotation}
\noindent  
``And what does this have to say about the original motivation for Kalai's conjectures, namely,
skepticism of fault-tolerant quantum computing? While our examples certainly do not prove anything about the
feasibility of fault-tolerant quantum computing, we do believe they point to the
difficulty of defining simple physically observable signatures of quantum computing that could be
said to be different from previously observed physics.''

\end {quotation}

Thought experiments are useful but they cannot be used on their own
without firm empirical support to describe ``previously observed physics.''

\subsection* {3. The fallacy that everything not formally prohibited by quantum physics is possible} 

Another quote from Flammia and Harrow's paper is:

\begin {quotation}
\noindent  
``Quantum error-correcting code states are
known to be robust to error (indeed, they are designed for this), but constructing them is a
significant challenge that skeptics may doubt on intuitive grounds, even if quantum mechanics
doesn't formally prohibit this.''

\end {quotation}

As a matter of fact, random quantum states that Flammia and Harrow discuss in the next quote
are not prohibited by quantum mechanics but are prohibited by a simple computational complexity argument on top
of the laws of quantum mechanics.

\subsection* {4. The fallacy of drawing wrong conclusions from the example of random quantum states}

  Our fourth item refers to the following quote from \cite {FlaHar13}:

\begin {quotation} 
\noindent  
``In this note, we attempted to argue that in fact
robust entanglement is not very hard to produce.
One intuitive explanation for this is related to the fact that random states are entangled
with overwhelming probability. There simply aren't that many product states out there, and even
the volume of separable states is a small fraction of the set of all possible density matrices.''

\end {quotation}

Random quantum states 
are good examples of states that are not prohibited by quantum physics but would be out of reach
even if quantum computers could be built. The reason
is that the volume of states that can be achieved by {\it efficient} computation is
a small fraction of all states.\footnote {The widely accepted {\it quantum
    extended Church-Turing thesis} (QECTT) asserts that all quantum processes in nature can be efficiently
  simulated by a standard quantum computer.}
In \cite {FlaHar13} random quantum states are used to give an intuitive explanation for why
robust entanglement is not very hard to produce. It is a fallacy to base such an argument on states
that unequivocally cannot be produced.

{\bf Remark:}  One could replace ``random states'' by ``pseudorandom states'' in the argument. However,
I am not aware of robust pseudorandom states
that have been empirically created. (The states created in \cite {Arut+19} have very
low fidelity and are also not robust.) 
In fact, an important ingredient of the author's general argument from \cite {Kal20,Kal18}
is that pseudorandom quantum states are themselves
Sure/Shor separators and, in addition,  are inherently non-robust.



\section {Three conjectures on physics and computation}
\label {pc}

We briefly discuss three conjectures on the computational power of physical systems which are related to
the debate on quantum computers.

\begin {itemize}

\item [1)]
  The {\it extended Church--Turing thesis} (ECTT) asserts that all computations in nature can be efficiently
carried out by a standard digital computer.

\end {itemize}

The term ``efficiently'' is asymptotic and asserts that the number of
computational steps required by the digital computer is at most polynomial in the input size. Given the
standard (mathematical) conjectures in computational complexity, quantum computers will violate the ECTT.
Pitowsky's paper \cite {Pit90} is one of the earliest papers to
study ECCT (there it is called ``the physical Church--Turing thesis).
Pitowsky attributes the extended Church--Turing thesis to Wolfram \cite {Wol85}.   

\begin {itemize}

\item [2)]
The {\it quantum extended Church--Turing thesis} (QECTT) asserts that all computations in nature can be efficiently
simulated by a standard quantum computer.

\end {itemize}
QECTT does not formally follow from quantum mechanics but it can be derived by
a heuristic argument based on quantum mechanics and computational complexity.
QECTT implies that random quantum states cannot be achieved, and
with some plausible mathematical computational complexity
conjectures, QECTT implies that NP-complete problems are beyond reach for computations in nature.

\begin {itemize}

\item [3)]

  The {\it NISQ=LDP conjecture} asserts that probability distributions
described by noisy intermediate scale
quantum computers can be well approximated by low degree polynomials \cite {Kal20}.

\end {itemize}

The QECTT as well as the stronger ECTT apply to computations by general physical systems when the number of
qubits tends to infinity. In contrast, the NISQ=LDP conjecture applies to computations
by general noisy quantum physical
systems, when the number of qubits is in the small and intermediate scale.
This conjecture
was stated (in a slightly different way) in \cite {Kal20} where the connection to various forms of the
extended Church--Turing thesis is discussed. In particular, \cite {Kal20}  provides an argument for the validity of
NISQ=LDP, as well as an argument for why this conjecture on the behavior of physical computational
devices in the intermediate scale implies the ECCT that deals with the
asymptotic behavior of physical computational 
devices.

Both QECTT and NISQ=LDP do not formally follow from quantum mechanics but are derived by heuristic arguments.
Because of their asymptotic ingredients, applying both QECTT and NISQ=LDP to specific claims is also heuristic.
For example: QECTT (heuristically) implies that we will never be able to compute the permanent
of a general 1000 by 1000 matrix. The NISQ=LDP conjecture (heuristically) implies that we will never be
able to perform on a NISQ computer 
a task that requires 10,000 years for a powerful classical supercomputer (as was claimed in \cite{Arut+19}).
The argument in \cite {Kal20} shows
how the NISQ=LDP conjecture implies the assertion that distance-5 surface codes are Sure/Shor separators. 

\begin{figure}[H]
\centering
\includegraphics[scale=0.5]{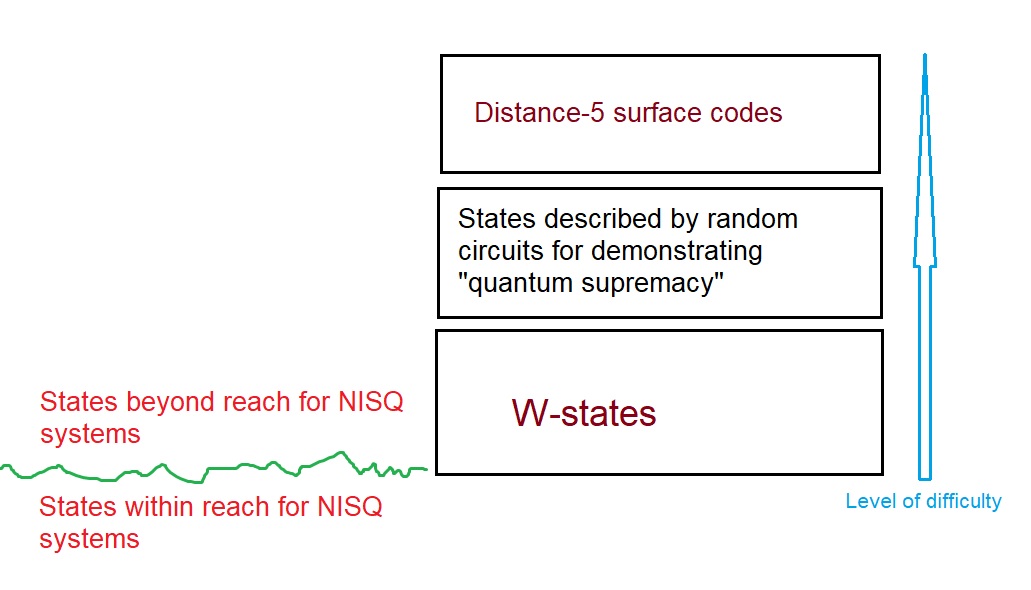}
\caption{{\it Creating $W_n$ states on NISQ computers would be easier than
  creating states required for ``quantum supremacy demonstration,'' and creating these states would be
  easier than creating distance-5 surface code.} }
\end{figure}

\section {Conclusion}

Flammia and Harrow conclude their article with the following statement.
\begin {quotation}
\noindent    
``Our note can be
thought of as showing that Conjecture C refers to a correlation measure that is high not only for
full-scale quantum computers, but even for the quantum equivalent of light bulbs, i.e. technology
which is non-trivial, but by no means complex.''

\end {quotation}

The fact of the matter is that $W$-states are indeed simple and can be seen as quantum equivalent of light
bulbs, but it is plausible that  
even these uncomplicated states %
cannot be achieved by noisy quantum computers
without quantum fault-tolerance, and there is no sufficient support to the claim that such states
(with exponential value of $K(\rho)$) can be seen in nature and
indeed represent ``previously observed physics.''
It is indeed a fascinating question raised by Flammia and Harrow's paper \cite {FlaHar13} whether
$W_n$ states 
(with exponential value of $K(\rho)$) can be seen in nature.

Creating high-fidelity $W_n$ states with twenty, thirty, fifty, and hundred
qubits could serve as useful benchmarks for NISQ computers toward the goal of creating the more
complicated states needed for quantum fault-tolerance.


\subsection *{Acknowledgments}
I am thankful to Steve Flammia, Aram Harrow, and Greg Kuperberg for helpful discussions. I am especially thankful to Steve for
bringing \cite {MM} to my attention. This work was supported by ERC advanced grant 834735.

\bibliographystyle{emss}

\end {document}